\documentclass[a4paper,11pt]{article}
\usepackage[utf8]{inputenc}
\usepackage{amsmath}
\usepackage{amssymb}
\usepackage{graphicx}
\newcommand{\Var}{\mathrm{Var}}

\begin{document}

\title{A simple model describing evolution of genomic GC content with random perturbations in asexually reproducing organisms}
\author{Jon Bohlin\textsuperscript{1,2,}\footnote{email:jon.bohlin@fhi.no}}

\maketitle

\textsuperscript{1}Division of Infection Control, Department of Methods Development and Analysis, Norwegian Institute of Public Health\par
\textsuperscript{2}Centre for Fertility and Health, Norwegian Institute of Public Health\par
Norwegian Institute of Public Health, P.O. Box 4404, Lovisenberggata 8, 0403 Oslo, Norway.\par

\begin{abstract}
A genome’s nucleotide composition can usually be summarized with (G)uanine + (C)ytosine (GC) or (A)denine + (T)hymine (AT) frequencies as GC\%=100\%-AT\%. A model is presented relating the evolution of genomic GC content over time to AT$\rightarrow$GC and GC$\rightarrow$AT mutation rates. By employing It\^o calculus it is shown that if mutation rates in asexually reproducing organisms are subject to random perturbations that can vary over time several implications follow. For instance, an extra Brownian motion term appears influencing nucleotide variability; the greater the variability of  the random perturbations on the mutation rates the stronger the impact of the Brownian motion term. Reducing the influence of the random perturbations, to limit fitness decreasing and deleterious mutations, will likely imply divesting resources to genomic repair systems. The stable mutation rates seen in many organisms could thus be an evolved strategy to reduce the influence of the Brownian motion term. Furthermore, if change to genomic GC content, \emph{i.e.} the GC content of variable sites or single nucleotide polymorphisms (SNPs), is just as likely to increase as to decrease, something that resembles knockout of repair enzymes and removal of selective pressures seen in evolutionary laboratory experiments, the species genome will likely decay unless infinite resources are available. These implications are solely a consequence of allowing random perturbations affect AT- and GC mutation rates and not obtainable using standard non-stochastic methodology. Finally, a connection between the model for genomic GC content evolution and the classical Luria-Delbrück mutation model is presented in a stochastic setting.
\end{abstract}

\section{Introduction}
The hereditary material of living organisms consist of double stranded deoxyribonucleic acid (DNA) molecules \cite{Watson1953}. The building blocks of the DNA molecule are the nucleotides Adenine (A), Guanine (G), Cytosine (C) and Thymine (T) \cite{Watson1953}. Across each strand Gs pair with Cs and Ts with As only \cite{Watson1953}. Within each strand the different nucleotides are stacked in no specific order connected via a sugar backbone \cite{Watson1953}. Virus can have genomes consisting of single or double stranded DNA or ribonucleic acid (RNA) \cite{Eigen1993}. RNA is similar to DNA but is more often single stranded and T is substituted for Uracil (U) \cite{Meyer2021}. Genomes consisting of double stranded DNA have approximately as many As as Ts and Gs as Cs on each strand \cite{Chargaff1951}. These relations were first observed by Erwin Chargaff \cite{Chargaff1951} and are therefore referred to as one of Chargaff's parity laws. Base composition in genomes with double stranded DNA can therefore be analyzed using either AT- or GC content as GC\%=100\%-AT\%. Base-pairing across the two DNA strands consists of three hydrogen bonds for G and C and two for A and T \cite{Watson1953}. More energy is in general required to stack and melt G and C bindings as compared to A and T \cite{Chen2015}. Methylation of C, by the addition of a methyl group, often results in deamination of C transforming it to T \cite{Bohlin2019b}. Deamination of methylated Cs is therefore implicated in an AT mutational bias observed for bacteria and archaea (prokaryotes) \cite{Hershberg2010}. Relaxation of selective pressures are hypothesized to increase AT content in microbial genomes due to the failure of repair enzymes to remove methylated cytosines \cite{Couce2017}. Indeed, both modelling and empirical investigations points to an approximately 2:1 relationship of respectively GC$\rightarrow$AT and AT$\rightarrow$GC mutation rates in prokaryotes \cite{Hershberg2010,Bohlin2019a}. DNA methylation also occurs in organisms with a cell nucleus (eukaryotes) and it is therefore reasonable to expect an AT mutational bias for these organisms as well \cite{Bohlin2019a}. However, as many larger, multi-cellular organisms reproduce sexually homologous recombination may obscure any such mutational AT bias observed for prokaryotes \cite{Bohlin2019b}. Homologous recombination seems instead to increase GC content in eukaryotes as a consequence of a process referred to as GC-biased gene conversion \cite{Rouselle2019}. Examination of AT- and GC mutation rates in prokaryotes points to an AT bias also in recombining microbes while AT $\rightarrow$ GC substitutions are more likely to be retained \cite{Bohlin2018, Bohlin2019a}. It should be noted that AT $\rightarrow$ GC mutations (and vice versa) are in the present work taken to mean all possible combinations of A and T to G and C (G and C to A and T).\par
For larger multi-cellular organisms the GC content is fairly stable between species although there can be local genomic differences, for instance CpG islands which are usually not present in prokaryotes \cite{Bohlin2019b}. For prokaryotes within-species genomic GC content is stable while GC content between different species can vary substantially, from 13.5\% GC to 75\% GC \cite{Bohlin2020}. This variation in genomic GC content appears to have both an environmental component as well as a phylogenetic one \cite{Foerstner2005,Reichenberger2015}. Environmental influence on genomic GC content in prokaryotes appears to be mediated, at least to some extent, as selective pressures \cite{Hildebrand2010}. In this respect, it is of interest to note that large drops in genomic GC content is wide-spread in prokaryotes \cite{Hershberg2010} while examples of substantial increase is practically non-existent as of now \cite{Ely2021}, although smaller increases are documented \cite{Raghavan2012, Reichenberger2015, Hershberg2015, Bohlin2017}.\par
A previous study \cite{Bohlin2020} modelling the evolution of GC content in microbes living in a stable symbiotic relationship with an eukaryotic host, usually insects, suggested that AT $\rightarrow$ GC mutation rates, and vice versa, may determine the symbiotic species fate early on in the organism's history providing mutation rates are approximately constant. Microbial symbionts often live in low density populations, are unable to perform homologous recombination and lack many DNA repair genes implying that mutations accumulate, often in a clock-like manner \cite{McCutcheon2012}. If the selective pressures are not strong enough to purge deleterious and fitness decreasing mutations from the symbiont's genome, it will decay \cite{Klasson2017}. This process is known as Muller's ratchet \cite{Moran1996}. For clonal, non-recombining symbionts, the evolutionary process of Muller's ratchet will always incur, the only question is when \cite{Moran1996}.\par
Since microbial symbionts live in a stable environment with their eukaryotic host it makes sense to model the random perturbations of their AT- and GC mutation rates as a Gaussian white noise multiplied by a constant, or a fixed parameter, to be estimated, as has previously been done \cite{Bohlin2020}. For asexually reproducing organisms in general, subjected to differing selective pressures over time, it is more natural to model random perturbations of AT/GC mutation rates as a Gaussian white noise multiplied by a function varying with respect to time. As such, the purpose of the present work is to expand on the previous study \cite{Bohlin2020}. That is, to model the evolution of genomic GC content as a consequence of AT $\rightarrow$ GC, as well as GC $\rightarrow$ AT, mutation rates subject to random perturbations $c(t)$ over time $t$, as opposed to a constant $c$, for all non-recombining, Chargaff parity law compliant organisms, regardless of which kingdom they belong to.

\section{Methods}

\subsection{The mathematical model}
\label{TheMathematicalModel}
The model presented in this study is an extension of models from previous work where details on how these formulas were derived can be found \cite{Bohlin2018,Bohlin2019a}. In particular, the present work is an extension of a model describing the evolution of genomic GC content as  a consequence of AT/GC mutation rates with random perturbations for microbial symbionts \cite{Bohlin2020}. As there are many similarities between present- and the previous work on microbial symbionts \cite{Bohlin2020} only the steps separating these models and details necessary for a complete comprehension are included.\par
First, $F_t(\omega)$ represents genomic GC content at time $t$ for all trajectories $\omega\in\Omega$ such that:

\begin{equation}
\label{DiscreteEquation}
F_{t+\Delta t} (\omega) - F_t (\omega) = \alpha F_t (\omega)\Delta t+\beta(1-F_t (\omega))\Delta t
\end{equation}

That is, the change in genomic GC content $F_{t+\Delta t} (\omega) - F_t (\omega)$ during time $\Delta t$ is a fraction multiplied with genomic GC- and AT content, respectively. Somewhat inaccurately this will be interpreted as the GC content of single nucleotide polymorphisms (SNPs/variable sites) in a species during time $\Delta t$. It will be shown later that there is a natural way of accurately extracting SNP GC/AT content from eq. (\ref{DiscreteEquation}). If it is assumed that as $\Delta t\rightarrow 0$ eq. (\ref{DiscreteEquation}) can be written as a differential equation:

\begin{equation}
\label{DeterministicEquation}
\frac{dF_t (\omega)}{dt} = \alpha F_t (\omega) + \beta (1-F_t (\omega)),
\end{equation}

GC- and AT mutation rates are described by, $\alpha F_t (\omega)$ and $\beta (1-F_t (\omega))$, respectively. The AT$\rightarrow$GC and GC$\rightarrow$AT mutation rates are subject to random perturbations that can vary with time multiplied by a function $c(t)$, \emph{i.e.} $\alpha=a+c(t)W_t (\omega)$ and $\beta=b+c(t)W_t (\omega)$. It is therefore assumed that $c(t)$ is a measurable function and that $W_t (\omega)$ is a Gaussian white noise \cite{Oksendal2003} with respect to all trajectories $\omega \in \Omega$. Furthermore, eq.  (\ref{DeterministicEquation}) is taken to belong to the probability space $(\Omega, \mathcal{F} _t, P)$ while $c(t)$ is an element of the measure space $(\mathbb{R}^+,\mathcal{G}, dt)$. $\mathcal{F}_t$ is the filtration of $\Omega$ with respect to each time $t\in\mathbb{R^+}$ (\emph{i.e.} $[0,\infty)$ of which $\mathcal{G}$ is a Borel algebra and $dt$ the corresponding Lebesgue measure), and $P$ is a probability (Lebesgue) measure on $\Omega$. The filtration $\mathcal{F} _t$ is interpreted as the evolutionary history of trajectories $\omega$ up to time $t$. Some further re-arrangements gives:

\begin{eqnarray}
\frac{dF_t (\omega)}{dt} &&= (a+c(t)W_t (\omega)) F_t (\omega) + (b+c(t)W_t (\omega)) (1-F_t (\omega)) \nonumber \\ \nonumber
						&&=aF_t (\omega)+c(t)W_t(\omega)F_t(\omega)+\\ \nonumber
						&&+b(1-F_t (\omega))+c(t)W_t (\omega) (1-F_t (\omega))\\ \nonumber
						&&=aF_t (\omega) + b(1-F_t (\omega))+c(t)W_t (\omega) \nonumber.
\end{eqnarray}

The equation:

\begin{equation}
\label{MainEquation}
\frac{dF_t (\omega)}{dt} =aF_t (\omega) + b(1-F_t (\omega))+c(t)W_t (\omega),
\end{equation}

can be written as a differential form as was explained in the previous study \cite{Bohlin2020}:

\begin{equation}
\label{MainEquationDifferential}
dF_t (\omega) =(aF_t (\omega) + b(1-F_t (\omega)))dt+c(t)dB_t (\omega).
\end{equation}

If $c(t)=c$, where $c$ is a constant, eq. (\ref{MainEquationDifferential}) will coincide with the model for microbial symbionts \cite{Bohlin2020} and so is an extension containing that model. The Brownian motion term can additionally be a $c$-scaled Brownian motion which was previously also shown \cite{Bohlin2020} to be a Brownian motion there termed $\hat{B_t}$. Using the It\^o formula \cite{Oksendal2003}:

\begin{equation}
\label{six}
dY_t (\omega) =\frac{\partial g}{\partial t} (t,F_t(\omega))dt+\frac{\partial g}{\partial t}(t, 	F_t(\omega))dF_t(\omega)+\frac{1}{2} \frac{\partial ^2g}{\partial x^2} (t,F_t (\omega))(dF_t (\omega))^2
\end{equation}

eq. (\ref{MainEquationDifferential}) can be given an explicit solution through the integrating factor $g(t,F_t (\omega))=Y_t (\omega)=e^{(-(a-b)t)} F_t (\omega)$:

Because $\frac{\partial ^2 g}{\partial x^2}(t,x)=0$ (see \cite{Oksendal2003} for details), the last term of eq. (\ref{six}) is equal to zero. As a result,
\begin{eqnarray}
dY_t (\omega)&&=\frac{\partial g}{\partial t} (t,F_t(\omega))dt+\frac{\partial g}{\partial x}(t, 	F_t(\omega))dF_t(\omega)\nonumber \\
&&=-(a-b)e^{(-(a-b)t)} F_t(\omega)dt+e^{(-(a-b)t)}dF_t (\omega)\nonumber \\
&&=-(a-b)e^{(-(a-b)t)} F_t(\omega)dt+ \nonumber \\
&&+e^{(-(a-b)t)}(((a-b)F_t (\omega) +b)dt+c(t)dB_t) \nonumber \\
&&=be^{(-(a-b)t)}dt+c(t)e^{(-(a-b)t)}dB_t. \nonumber
\end{eqnarray}

$F_t(\omega)$ can then be give an explicit formula:

$$d(e^{(-(a-b)t)} F_t (\omega))=be^{(-(a-b)t)}dt+c(t)e^{(-(a-b)t)}dB_t$$
which, by assuming $s\in [0,t]$, gives
$$e^{(-(a-b)t)} F_t (\omega)-F_0(\omega)=\int_0 ^t be^{(-(a-b)s)}ds + \int_0 ^t c(s)e^{(-(a-b)s)} dB_s,$$
and therefore the integral

\begin{equation}
\label{FinalIntegral}
F_t (\omega)=F_0(\omega) e^{(a-b)t}+\int_0 ^t be^{(a-b)(t-s)}ds + \int_0 ^t c(s)e^{(a-b)(t-s)}dB_s.
\end{equation}

An analogous argument to the previous study \cite{Bohlin2020} gives an explicit formula for the finite variation term $\int_0 ^t be^{(a-b)(t-s)}ds$. The quadratic variation Brownian motion term $\int_0 ^t c(s)e^{(a-b)(t-s)}dB_s$ however must be approximated numerically, not least due to the unspecified measurable function $c(t)$:

\begin{equation}
\label{FinalFormula}
F_t (\omega)=-\frac{b}{(a-b)}+(F_0(\omega) +\frac{b}{(a-b)})e^{(a-b)t}+\int_0 ^t c(s)e^{(a-b)(t-s)}dB_s
\end{equation}
which is subject to the constraints $t\in[0,\infty)$ and $0<F_t (\omega)<1$. The integration constant $c_0$ is just included in $F_0$. It should be noted that for $F_0=0$,
\begin{equation}
\label{ExpectanceForumla}
\mathbb{E}(F_t(\omega))=\frac{b}{(a-b)}(e^{(a-b)t}-1).
\end{equation}
Since the Brownian motion term vanishes as a consequence of the expectation operator $\mathbb{E}$ (see p. 30 of \cite{Oksendal2003}), the solution to (\ref{FinalFormula}) when $t=x$ is exactly the model for SNP GC content with respect to core genome GC content $x$ described in previous work \cite{Bohlin2018, Bohlin2019a}. Moreover, this means that it is not necessary to calculate the Brownian motion term when estimating the AT- and GC mutation rate parameters $a$ and $b$.

The variance is given by $\Var(F_t (\omega))=\mathbb{E}((F_t(\omega)-\mathbb{E}(F_t(\omega))^2)$, which can be solved by setting:
$$A:=F_0(\omega) e^{(a-b)t}+\frac{b}{(a-b)}(e^{(a-b)t}-1)$$ and $$B:=\int_0 ^t c(s)e^{(a-b)(t-s)}dB_s$$ which gives:
\begin{eqnarray}
\Var(F_t (\omega))&&=\mathbb{E}((F_t(\omega)-\mathbb{E}(F_t(\omega))^2 \nonumber \\
	&&=\mathbb{E}((A+B)^2-2(A+B)A+A^2) \nonumber\\
	&&=\mathbb{E}(A^2+2AB+B^2-2A^2-2AB+A^2) \nonumber\\
	&&=\mathbb{E}(B^2)=\mathbb{E}((\int_0 ^t c(s)e^{(a-b)(t-s)}dB_s)^2). \nonumber
\end{eqnarray}
Applying the Itô isometry (see pg. 26 \cite{Oksendal2003}):
$$\mathbb{E}((\int_0 ^t c(s)e^{(a-b)(t-s)}dB_s)^2)=\mathbb{E}(\int_0 ^t (c(s)e^{(a-b)(t-s)})^2 ds)$$
$$=\int_0 ^t c(s)^2 e^{2(a-b)(t-s)} ds.$$
The equation cannot be given an analytic representation due to the unspecified function $c(t)$ but it is clear that the integral $\int_0 ^t c(s)^2 e^{2(a-b)(t-s)} ds \rightarrow \infty$ as $c(s)$ increases for $s\rightarrow t$. As was argued in the previous study on microbial symbionts \cite{Bohlin2020}, it will be assumed henceforth that $(a-b)<0$ where $a$ and $b$ are respectively the AT$\rightarrow$GC and GC$\rightarrow$AT mutation rate parameters that can be estimated (see \cite{Bohlin2018, Bohlin2019a}) or given. The previous study \cite{Bohlin2020} also showed that $a$ and $b$ could be considered as unspecified measurable functions. However, as constant mutation rates are not uncommon \cite{Lynch2010} $a$ and $b$ will henceforth be regarded as constants, or parameters to be estimated, to avoid unnecessary complication of interpretation, formulation and derivation of the model.

\subsection{The Brownian motion term}
The term:
\begin{equation}
\label{BrownianIntegral}
\int_0 ^t c(s)e^{(a-b)(t-s)}dB_s,
\end{equation}
depends on the parameters $a$ and $b$ as well as on the duration of the time period. Since it is assumed that $(a-b)< 0$, $c(s)e^{(a-b)(t-s)}\rightarrow c(s)$ for $s\rightarrow t$. The term can be written as:

\begin{equation}
\label{numLocMartingale}
\int_0 ^t c(s)e^{(a-b)(t-s)}dB_s=\lim_{\Delta s_i \rightarrow 0} \sum_{s_0} ^{s_N} c(s_i)e^{(a-b)(t-s_i)}(W_{s_{i+1}} (\omega) - W_{s_i}(\omega))\Delta s_i,
\end{equation}
where $W_s (\omega)$ is scaled white noise with mean $\mu=0$, variance $\sigma ^2 =1$ (see \cite{Bohlin2018}), $\Delta s_i =s_{i+1}-s_i$, and $s_0=0,\ldots,s_i=t_i,\ldots,s_N=t$. The right hand side term of eq. (\ref{numLocMartingale}) can be calculated manually by inserting values for each value $s_i$ and mutation parameters $a$ and $b$. Each Gaussian white noise $W_{s_i}(\omega)$ can be sampled from a normal distribution.

\subsection{The Girsanov Transform}
\label{TheGirsanovTransform}
The Girsanov transform implies that eq. (\ref{FinalFormula}), if an appropriate transform exists, can be considered as a Brownian motion. In other words, given the appropriate transform, genomic GC content can be seen to be just as likely to increase as to decrease. This is a scenario that can arise when mismatch and repair enzymes are knocked out and the species is subjected to reduced selective pressures which is sometimes recreated in laboratory settings such as the long term evolutionary experiment (LTEE) \cite{Couce2017}. The random perturbations of AT/GC mutation rates are represented here as Gaussian white noise multiplied by an unspecified deterministic measurable function $c(t)$. Together with the Gaussian white noise term $W_t(\omega)$, the function $c(t)$ accounts for environmental influences and species-specific traits perturbing the AT/GC mutation rate variances over time $t$. Since eq. (\ref{MainEquationDifferential}) is a differential equation the Gaussian white noise should turn into Brownian motion if the SNP GC content is to be modelled as being completely unbiased with regards to both selective pressures and mismatch and repair enzymes. If it is assumed that $t\in [0,T]$ for a fixed time $T$, then:

\begin{equation}
\label{GirsanovEquation}
dF_t (\omega) =((a-b)F_t (\omega) + b)dt+c(t)dB_t (\omega).
\end{equation}

To see that the Girsanov theorem applies to $F_t(\omega)$ recall from section \ref{TheMathematicalModel} that for $Y_t (\omega) = e^{(-(a-b)t)} F_t (\omega)$:

\begin{equation}
dY_t = be^{(-(a-b)t)}dt+c(t)e^{(-(a-b)t)}dB_t
\end{equation}

Since $F_t (\omega)$ is a semimartingale, the Girsanov II theorem (pg. 167, \cite{Oksendal2003}) can be used. Let

$$ c(t)e^{(-(a-b)t)}u(t,\omega)=be^{(-(a-b)t)}-\alpha(t,\omega) $$

and thus:

$$ u(t,\omega)=c^{-1}(t)(b-\alpha(t,\omega)e^{(a-b)t}) $$

$\alpha(t,\omega)$ can then be set to zero so that:

$$u(t,\omega)=c^{-1}(t)b$$

which means that $c(t)$ in eq. (\ref{MainEquationDifferential}) is required to have an inverse function $c^{-1}(t)$. Let

$$ M_t = \exp{\Big(-\int_0 ^t u(s,\omega)dB_s - \frac{1}{2}\int_0 ^t u^2 (s,\omega)ds\Big)}$$

and set
$$dQ(\omega) = M_t (\omega)dP(\omega)$$
with respect to the filtration $\mathcal{F}_t$, assume that the Novikov condition (see pg. 165 \cite{Oksendal2003}) holds:
$$E[\exp{\Big(\frac{1}{2} \int_0 ^t u^2(s,\omega)ds\Big)}]<\infty$$ so that the Radon-Nikodym derivative $M_t$ is a martingale. Then $Q$ is a probability measure with respect to the filtration $\mathcal{F}_t$ and

$$\hat{B_t}(\omega)=\int_0 ^t u(s,\omega)ds+B_t (\omega)$$

is a Brownian motion with regards to the measure $Q$ and so is $Y_t$, according to the Girsanov theorem:
$$ dY_t (\omega) = c(t)d\hat{B_t}(\omega)$$
Since $Y_t(\omega)=e^{(-(a-b)t)} F_t (\omega)$ it is clear that $F_t(\omega)$ is also a Brownian motion with respect to $Q$. Hence, given the right transform $F_t (\omega)$ is just as likely to increase as to decrease. If the transformed $F_t (\omega)$ is considered to represent a culture in an LTEE-type experiment \cite{Lang2013, Miller2011, Couce2017}, where mismatch and repair genes are knocked out and selective pressures are assumed to be at a minimum, the function $c(t)$ must have an inverse, \emph{i.e.} $c(t)$ must be either monotonically increasing or decreasing.

\section{Results and Discussion}
\subsection{Allowing random perturbations to vary as a function of time}
By allowing the random perturbation of AT/GC mutation rates to vary according to a function $c(t)$ a number of differences can be noted as compared to the previous model \cite{Bohlin2020} where $c(t)=c$. Indeed, while asexually reproducing organisms still will be subject to the process of Muller's ratchet \cite{Moran1996} this can no longer be predicted to the same extent previously described for microbial symbionts \cite{Bohlin2020}. In other words, as the random perturbations of the AT/GC mutation rates vary according to a function $c(t)$, with respect to time $t$, it is conceivable that selective pressures could completely alter the evolutionary trajectory of the organism. One of the surprising results of the model describing evolution of genomic GC content in microbial symbionts \cite{Bohlin2020} was that the fate of the symbiont could have been set when it entered it's symbiotic relationship possibly millions of years before an extinction event. Allowing the random perturbations of the AT/GC mutation rates to vary according to a function with regards to time $t$, as opposed to being fixed, the genomic GC content of the organism can vary as time progresses. Hence, selective pressures and genetic drift could radically change the evolutionary trajectory of the organism.

\subsection{Comparison of mathematical models}
As mentioned above, the present work is based on a previous study \cite{Bohlin2020} where the aim was to model the evolution of genomic GC content, as a consequence of AT/GC mutation rates, in microbial symbionts over time $t$, \emph{i.e.}:

$$F_{t+\Delta t} (\omega) - F_t (\omega) = \alpha F_t (\omega)\Delta t+\beta(1-F_t (\omega))\Delta t$$

which gave the differential equation:

$$\frac{dF_t (\omega)}{dt} = \alpha F_t (\omega) + \beta (1-F_t (\omega))$$

when $\Delta t\rightarrow 0$.\par
In other words, the change in the GC content of SNPs, or variable sites, over time $t$ is a fraction $\alpha$ multiplied by genomic GC content $F_t(\omega)$ in addition to a fraction $\beta$ multiplied by genomic AT content ($1-F_t(\omega)$, according to Chargaff's parity rules \cite{Chargaff1951}). Thus, $\alpha$ and $\beta$ designated the fraction of genomic GC content mutating to GC (\emph{i.e.} A\textgreater C ,A\textgreater G ,T\textgreater C ,T\textgreater G or G\textgreater C ,C\textgreater G) and to AT (C\textgreater A, G\textgreater A, C\textgreater T, G\textgreater T, A\textgreater T, T\textgreater A), respectively.

The model also allowed for random perturbations of SNP AT- and GC mutation rates, \emph{i.e.} $\alpha=a+cW_t (\omega)$ and $\beta=b+cW_t (\omega)$ where $W_t (\omega)$ is Gaussian white noise with respect to all trajectories $\omega \in \Omega$. After some rearrangement:

$$\frac{dF_t(\omega)}{dt}=aF_t (\omega) + b(1-F_t (\omega))+cW_t (\omega)$$

Since it was shown \cite{Bohlin2020} that the scaled volatility of Brownian motion is also a Brownian motion the equation could be solved analytically. It was also shown that the mutation rates $a$ and $b$ could be described as measurable functions $a(t)$ and $b(t)$, respectively. However, as $a(t)$ and $b(t)$ are unspecified analytic solutions would be impossible except for trivial cases of $a(t)$ and $b(t)$.\par
The present work is not limited to the genomes of microbial symbionts, which tend to live in a stable relationship with their hosts, but to all asexually reproducing organisms regardless of which kingdom they belong to, including virus. It is still assumed that the genomes of the organisms considered comply with Chargaff's parity laws \cite{Chargaff1951}. In particular, it is assumed that \% G is approximately equal to \% C and that \% A is similar to \% T on each strand in the genomes of the organisms considered. Although Chargaff's parity laws were stated for most organisms with double stranded DNA genomes they also apply to many viruses with single stranded RNA genomes \cite{Bohlin2019b}. Indeed, the pathogen responsible for the currently ongoing Covid-19 pandemic, SARS-CoV-2, has a single strand, positive sense, RNA genome that obeys Chargaff's parity rule with approximately 38\% GC \cite{Zhou2020,Wu2020}. Nevertheless, increasing the generality of the previous model to the genomes of all asexually reproducing organisms implies that the assumption of constantly scaled random perturbing AT/GC mutation rates, which could be justified in a stable host-symbiont relationship, is no longer tenable. It is now therefore assumed that the random perturbations to AT/GC mutation rates vary according to a Gaussian white noise multiplied by a measurable function $c(t)$ with respect to time $t$ as deduced in section \ref{TheMathematicalModel}, eq. (\ref{FinalFormula}): 
$$\frac{dF_t (\omega)}{dt} =aF_t (\omega) + b(1-F_t (\omega))+c(t)W_t (\omega)$$
If restrictions are imposed on the function $c(t)$ eq. (\ref{FinalFormula}) can be applicable to the Girsanov transform \cite{Oksendal2003}. SNP GC content is then just as likely to increase as to decrease with respect to the measure resulting from the Girsanov transform. In other words, it can then be guaranteed that $F_t(\omega)$ is a Brownian motion relative to a measure $Q(\omega)$. SNP GC content then follows a completely random path $\omega\in\Omega$ according to the law of Brownian motion. Due to influences from positive- and negative selection and mismatch and repair systems this is typically not observed outside laboratories \cite{Couce2017}.\par

\begin{figure}
\centering
\includegraphics[scale=0.63,bb = 0 0 7.8in 3.01in]{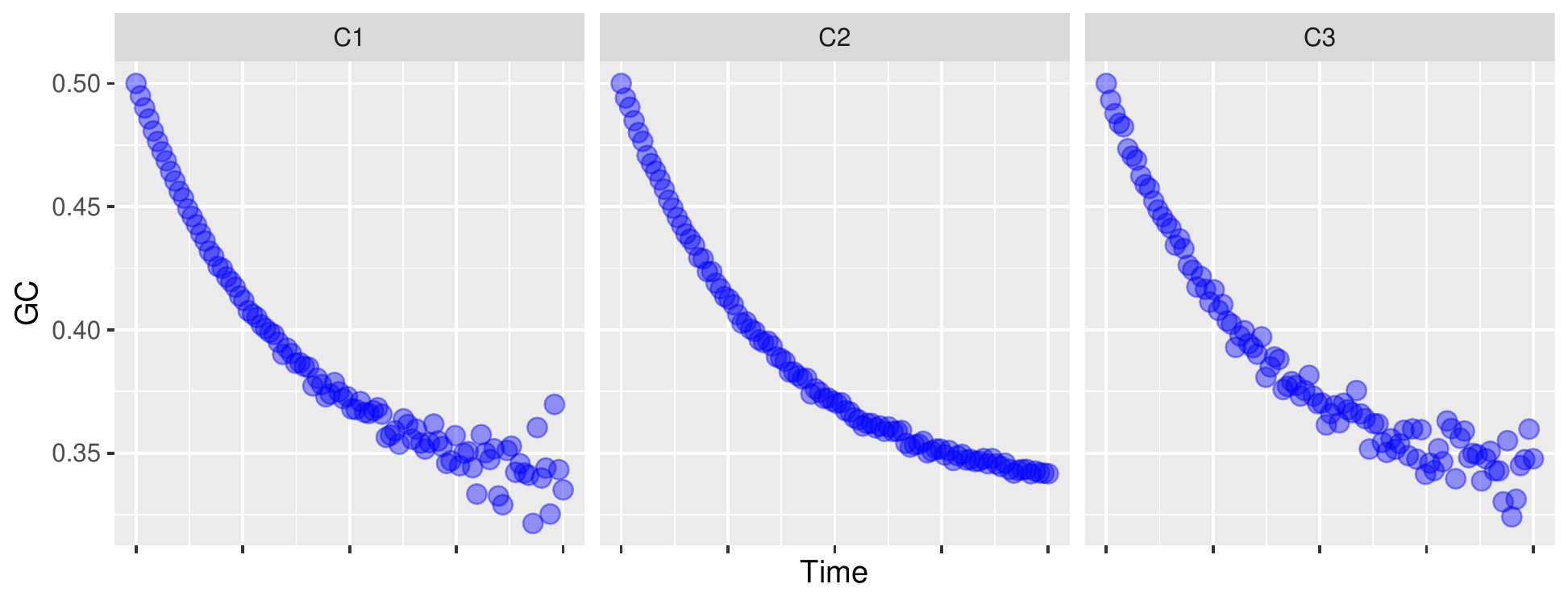}

\centering
\caption{\footnotesize The figure demonstrates three different (C1-C3) evolutionary scenarios for the model describing the evolution of genomic GC content (vertical axis) of a species over time $t$ (horizontal axis) as a consequence of AT/GC mutation rates. All parameters are the same for all scenarios (\emph{i.e.} $a=-2$, $b=1$, $F_0 =0.5$ and $T=1$) except for the function $c(t)$ that determines the influence of the random perturbations of the AT/GC mutation rates. Panel C1: $c(t)=\sqrt{t}$, Panel C2: $c(t)=T-t$, Panel C3: $c(t)= e^{{(T-t)^2}}$.}
\label{fig:1}
\end{figure}

\subsection{The evolutionary dynamics of $c(t)$}
AT/GC mutation rates are allowed to have random perturbation varying according to a function $c(t)$ with respect to time $t$, \emph{i.e.} $\alpha=a+c(t)W_t (\omega)$ and $\beta=b+c(t)W_t (\omega)$ where $W_t (\omega)$ is a Gaussian white noise with respect to every trajectory $\omega \in \Omega$. This has implications for the resulting Brownian motion term eq.  (\ref{BrownianIntegral}):
$$\int_0 ^t c(s)e^{(a-b)(t-s)}dB_s$$
Indeed, increasing $c(t)$ leads to greater variance in genomic GC content $F_t(\omega)$ as $s\rightarrow t$ while decreasing $c(t)$ reduces it (see Figure \ref{fig:1}). Increased variance of $F_t(\omega)$ can be interpreted as increasing the intrinsic genomic base composition variation. However, increasing the genomic base composition variance (\emph{i.e.} genomic GC content variance) may also lead to accumulation of fitness decreasing, and even deleterious, mutations \cite{Balbi2007,Bohlin2017}. Hence, increased variance with regards to perturbation of AT- and GC mutation rates highlights the trade-off between more genetic diversity versus an increased chance of accumulating fitness decreasing and deleterious mutations and subsequent extinction. For decreasing $c(t)$ approaching $0$ variation in base composition is reduced, something that is likely to happen in a scenario of increasing purifying selection \cite{Agashe2014,Bohlin2017}. Organisms living in large populations \cite{Lynch2016} or highly adapted to their environments \cite{Bohlin2017} are therefore more likely to have lower $c(t)$ values than organisms subjected to more diverse and changing environments. It has previously been shown that nucleotide diversity in microbes increases with AT content \cite{Bohlin2017}. That is, GC rich bacteria tend to have a more homogeneous nucleotide composition while AT rich bacteria have genomes comparably more heterogeneous in terms of nucleotide composition \cite{Bohlin2009}. The underlying reasons for this decreasing nucleotide diversity gradient, from AT rich to GC rich genomes, is not known but years of research has shed some light on the issue. For instance, GC rich bacteria are mostly found in soil, with excess nitrogen \cite{Agashe2014}. Soil bacteria have large genomes and are often capable of metabolising a wide range of compounds \cite{Reichenberger2015}. AT-rich bacteria, on the other hand, are often symbionts \cite{Moran2014} and pathogens \cite{Weinert2017} with reduced genome sizes. As mutations in prokaryotes are universally biased towards AT \cite{Hershberg2010} loss of proof reading enzymes as well as mismatch and repair genes, both hallmarks of relaxed selection, could be driving the greater genetic diversity found in these bacteria \cite{RochaFeil2010}. It has also been shown that the energetics of stacked A/T and G/C nucleotides is important in establishing genomic base composition \cite{Chen2015}. Between-strand binding of A to T requires only two hydrogen bonds as compared to the three hydrogen bonds required for G to C bindings \cite{Watson1953}. Guanine and cytosine also require the availability of more nitrogen \cite{Seward2016}. Within eukaryotic genomes, GC content vary considerably more than within prokaryotic genomes \cite{Bohlin2019b}. The genomes of most eukaryotic species have a considerably lower fraction of gene-coding DNA allowing for greater variation in GC content and thus a more varied $c(t)$ \cite{Bohlin2019b}. Viral genomes often mimic their hosts with respect to GC content and since their genomes are small, with a high fraction of coding genes, genomic GC content is typically far more stable than that observed for eukaryotic genomes and often comparable to prokaryotes \cite{Bohlin2019b}. Thus it would expected that $c(t)$ is lower for both virus and prokaryotes throughout their genomes as compared to most eukaryotes.
\subsection{Evolutionary implications}
The selective pressures a species is subjected to can, to some extent, be reflected by the trends of the measurable function $c(t)$. On one hand, $c(t)$ should be as close to zero as possible to avoid excessive hitch-hiking of fitness decreasing or deleterious mutations \cite{Balbi2007}. However, it will likely require several trade-off's for species' to reduce the random variation of AT- and GC mutation rates and thus keep $c(t)$ low; considerable resources must likely be divested to several genomic processes for $c(t)\rightarrow 0$. On the other hand, a changing environment may require species' to adapt rapidly implying the availability of an increased number of genotypes. A greater number of genetic variants require that the mutation rates reach a level that maximizes chance for survival of the species \cite{Couce2019}. At the same time, if mutation rates increase to such an extent that deleterious mutations cannot be avoided the evolutionary process of Muller's ratchet will ensue \cite{Moran1996}.\par
Whether mutations increase or decrease genomic GC content depends on the environment and the selective pressures operating on the species' genomes. Some environments could favour energetically affordable A/T nucleotides while others might require the more costly G/C nucleotides \cite{Chen2015}. In addition, phylogeny will also influence the selection of A/T or G/C nucleotides \cite{Reichenberger2015} as a consequence of the mismatch repair system and/or proof reading enzymes \cite{Bohlin2017, Couce2017, Lind2008}.\par
While there are many examples of microbial genomes becoming more AT rich \cite{Agashe2014} there are so few examples of genomes becoming more GC rich that it was recently suggested that it may not happen at all \cite{Ely2021}. Some examples have however been observed in the microbial world although the increase is miniscule \cite{Bohlin2017, Hildebrand2010, Bobay2017}. It is not completely resolved whether the few examples of microbial genomes becoming more GC rich is tied to recombination \cite{Lassalle2015}, which seems to be the case for recombining eukaryotes, or selection \cite{Hildebrand2010, Raghavan2012, Bohlin2019b}.\par
Muller's ratchet is, in the model discussed here, linked to the Brownian motion term and the measurable function $c(t)$ overwhelming the random perturbations of the AT/GC mutations rates. Hence, restrictions on $c(t)$ will necessarily influence the species' trajectory of genomic GC content and thus it's life course. It is interesting to note that allowing for random perturbation of AT/GC mutation rates introduces a term that will unequivocally lead to greater variability of genomic base composition (see eq. (\ref{FinalIntegral})); the larger the perturbations the greater the impact of the Brownian motion term on genomic variability. Put differently, genomic base composition can be modified by varying the random perturbations of the AT/GC mutation rates. It is therefore not unlikely that this is one of the reasons that mutation rates are occasionally found to be remarkably stable over a diverse set of organisms and, in particular, negatively associated with population size \cite{Giovanni2005, Lynch2010, Lynch2016}.\par
Assuming that eq. (\ref{FinalFormula}) is a Brownian motion, in the sense that it is just as likely that GC content will increase as it will decrease, it is required by the Girsanov theorem, as seen in section \ref{TheGirsanovTransform}, that $c(t)$ has an inverse function meaning that $c(t)$ must either be increasing or decreasing. All being equal, an increasing $c(t)$ will more likely lead to Muller's ratchet, and thus subsequent extinction, due to the accumulation of fitness decreasing and deleterious mutations. In other words, if it can be argued that $c(t)$ cannot be decreasing (\emph{i.e.} resources are finite), increasing random perturbations affecting AT- and GC mutation rates will necessarily lead to decay. Interestingly, this has been demonstrated in a laboratory experiment and described in a recent study based on the LTEE \cite{Couce2017}. \par
\subsection{Evolution of genomic GC content and the Luria-Delbrück mutation model}
\label{LDModel}

By setting $\beta=0$ in

$$\frac{dF_t (\omega)}{dt} = \alpha F_t (\omega) + \beta (1-F_t (\omega))$$

eq. (\ref{MainEquation}) can be interpreted as a simple model for stochastic population growth (pg. 65, \cite{Oksendal2003}):

$$\frac{dP_t (\omega)}{dt} = \pi P_t (\omega)$$

where $\pi = k + p(t)W_t(\omega)$, $k$ is the growth parameter to be estimated and $p(t)$ is a measurable function with respect to time $t$. By using the It\^o formula, along the lines described by \cite{Oksendal2003} in pg. 66, to solve for $P_t(\omega)$, and multiplying with a parameter $\mu$, taken to be a parameter designating mutations per unit time, a simple model for calculating the number of mutations in a study population is obtained. This is a stochastic version of the model presented in the classical Luria-Delbr\"uck fluctuation experiment \cite{Luria1943}, \emph{i.e.}

$$M_t(\omega)=P_t(\omega) \mu = P_0\mu \exp{((k-\frac{p^2 (t)}{2})t + p(t) B_t(\omega))}$$

\begin{figure}%
\centering
\includegraphics[scale=0.63,bb = 0 0 7.8in 3.01in]{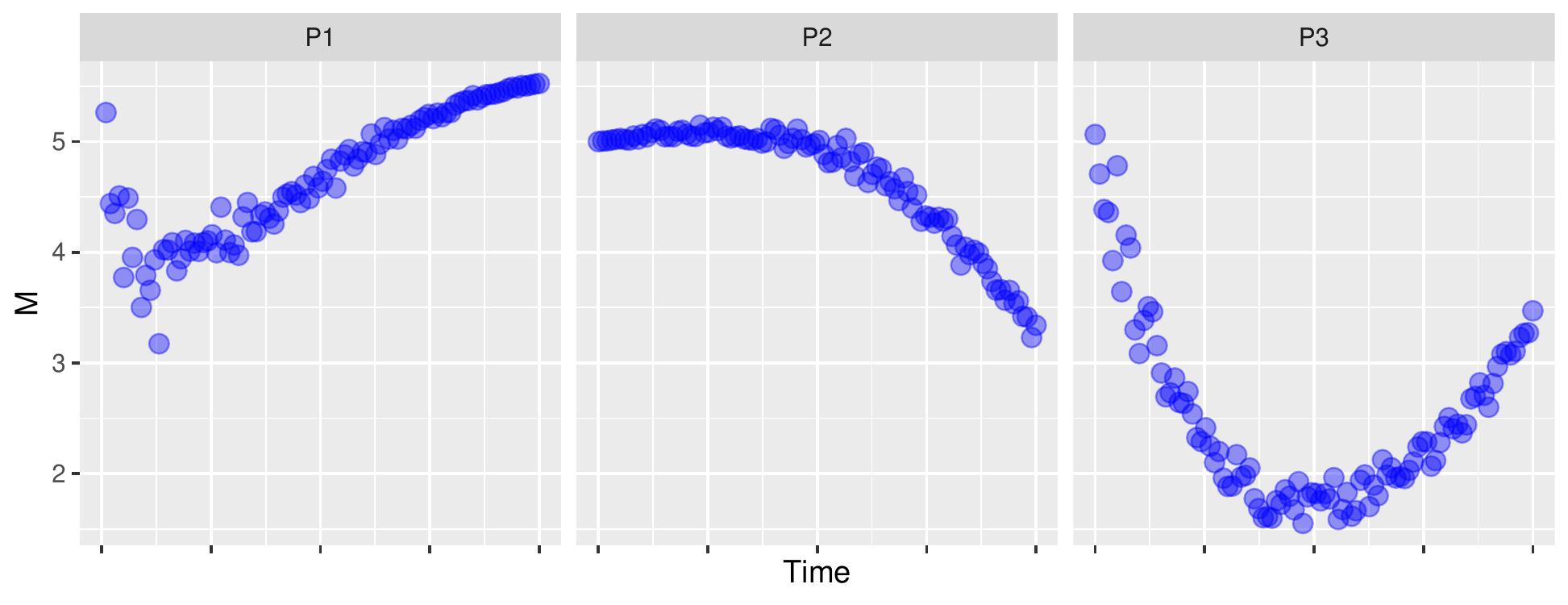}
\centering
\caption{\footnotesize The figure demonstrates three different (P1-P3) scenarios for the stochastic Luria-Delbrück model describing average number of mutations $M_t(\omega)$ (vertical axis) in a population at time $t$ (horizontal axis). All parameters are the same for all scenarios ($P_0 =1$, $\mu=5$, $k=1$ and $T=1$) except for the function $p(t)$ that determines the influence of random perturbation on growth rate. Panel P1: $p(t)=\ln{(t)}$, Panel P2: $p(t)=t$, Panel P3: $p(t)= e^{(T-t^2)}$.}
\label{fig:2}
\end{figure}

Hence, in this simple model the number of mutations in the study population is equal to the population size $P_t(\omega)$ multiplied with the mutation rate $\mu$. It can be seen from the above equations that as $k$ approaches $\frac{p^2 (t)}{2}$ a randomly fluctuating population will increasingly influence mutation rates. In other words, as the size of the population declines stochastic effects become more dominant \cite{Long2018} resulting not only in increased genetic variation (see Figure \ref{fig:2}) but also the danger of decay due to accumulation of fitness decreasing and deleterious mutations \cite{Lynch2016}.\par

$M_t(\omega)$ represents the total number of mutations in a population $P_t(\omega)$,  typically represented by mutations in one genome. If we assume that also $M_t(\omega)$ complies with Chargaff's parity rules (see \cite{Bohlin2018}) we can write:
$$ M_t(\omega) = M_t ^{AT} (\omega) + M_t ^{GC} (\omega) $$

where $M_t ^{AT} (\omega)$ and $M_t ^{GC} (\omega)$ represents the number of A+T and G+C mutations, respectively. Eq. (\ref{MainEquation}) gives the fraction of GC mutations at time $t$ and can be interpreted as the GC content of SNPs at that time. By multiplying eq. (\ref{MainEquation}) with the genome size $g$ the number of GC mutations is given:

$$g \cdot \Big\vert \frac{dF_t (\omega)}{dt} = \alpha F_t (\omega) + \beta (1-F_t (\omega)) $$

However, $g \frac{dF_t (\omega)}{dt} $ will only be equal to $M_t (\omega)$ if $M_t (\omega) = M_t ^{GC} (\omega) $. If $M_t (\omega) = M_t ^{AT} (\omega) $, $g \frac{dF_t (\omega)}{dt} $ will be negative which is impossible to reconcile with the fact that $M_t(\omega)\geq 0$. But since the right hand side of eq. (\ref{MainEquation}), multiplied with $g$, can be written as:

\begin{equation}
\label{GCMutations}
g \cdot \Big\vert \alpha F_t (\omega) + \beta (1-F_t (\omega))
\end{equation}

it is clear from eq. (\ref{GCMutations}) that $M_t ^{GC} (\omega)$ and $M_t ^{AT} (\omega)$ respectively correspond to $|g\alpha F_t (\omega)|$ and $|g\beta (1-F_t (\omega))|$. The triangle inequality with respect to eq. (\ref{GCMutations}) therefore gives:

$$M_t(\omega) = M_t ^{AT} (\omega) + M_t ^{GC} (\omega) = g\cdot(|\alpha F_t (\omega)| + |\beta (1-F_t (\omega))|)$$

which is the number of mutations at time $t$.

A model for mutation accumulation can be written as the Lebesgue integral:

$$\int_0 ^t M_s(\omega) ds$$

or:

$$\int_0 ^t g\cdot(|\alpha F_s (\omega)| + |\beta (1-F_s (\omega))|) ds$$

\section{Conclusions}
The presented work has been concerned with modelling the evolution of genomic GC content, as a consequence of AT/GC mutation rates, in asexually reproducing organisms subject to Chargaff's parity laws \cite{Chargaff1951}. It is an extension of a previous study modelling genomic GC content in microbial symbionts allowing for random perturbations of AT- and GC mutation rates by the use of It\^o calculus \cite{Oksendal2003}. In that study \cite{Bohlin2020}, it was shown that a symbiont's life course could be determined when it entered into a relationship with it's host. The present model does not allow for such a conclusion in general as the organisms modelled can both be diverse and live in very different environments. The fate of these organisms is thus linked to the function $c(t)$ regulating the influence of the random perturbations on AT/GC mutation rates that was taken to be constant for microbial symbionts. Increasing $c(t)$ will increase base composition diversity but also result in genetic hitch-hiking of fitness decreasing and deleterious mutations \cite{Balbi2007}. Keeping $c(t)$ low will likely reduce genetic variation but require the divestment of increasing resources to mismatch and repair systems as $c(t)\rightarrow 0$. Interestingly, eq. (\ref{FinalFormula}) implies that increasing the variability of the random perturbations of the AT- and GC mutation rates impacts genomic GC content through the Brownian motion term. It is not uncommon to encounter stable mutation rates in different organisms \cite{Lynch2010}. Indeed, the presented model (eq. (\ref{FinalFormula})) suggests that this could be resulting from selection to diminish variability in genomic GC content as described by the Brownian motion term.\par
Laboratory based evolutionary experiments \cite{Miller2011, Lang2013, Couce2017} often arrange conditions such that the selective forces subjected to the species' studied are as low as possible. Furthermore, recombination related genes as well as mutation repair enzymes are often knocked out \cite{Couce2017} reducing bias considerably with regards to AT$\rightarrow$GC and GC$\rightarrow$AT mutation rates. The model presented suggests that if genomic GC content is just as likely to increase as to decrease $c(t)$ must either be monotonically increasing- or decreasing due to constraints resulting from the Girsanov transform. Hence, if not resources are limitless $c(t)$ will increase eventually leading to genomic disintegration, as described by Muller's ratchet \cite{Moran1996}, something that has been demonstrated experimentally \cite{Couce2017}.\par
Finally, it is shown that there exists an intimate relationship with the model presented here and the classical Luria-Delbrück model for general mutations \cite{Luria1943}. Indeed, disregarding AT mutation rates by setting $\beta =0$ in eq. (\ref{MainEquation}) gives a model for stochastic population growth $P_t(\omega)$ \cite{Oksendal2003} which multiplied with a mutation rate $\mu$ gives a stochastic Luria-Delbrück model for the number of mutations $M_t(\omega)$. After some re-arrangements, it is shown in section \ref{LDModel} that eq. (\ref{MainEquation}) is related to $M_t(\omega)$.

\section*{Acknowledgements}
The figures were made with the free statistical software R \cite{R2021} and the \emph{ggplot2} package \cite{Wickham2016}.

\end{document}